# A Novel Hybrid Opportunistic Scalable Energy Efficient Routing Design For Low Power, Lossy Wireless Sensor Networks

Jayavignesh Thyagarajan and Suganthi K

School of Electronics Engineering, Vellore Institute of Technology,
Chennai Campus, India

*ABSTRACT*

*Opportunistic Routing (OR) scheme increases the transmission reliability despite the lossy wireless radio links by exploiting the broadcast nature of the wireless medium. However, OR schemes in low power Wireless Sensor Network (WSN) leads to energy drain in constrained sensor nodes due to constant overhearing, periodic beaconing for Neighbourhood Management (NM) and increase in packet header length to append priority wise sorted Forwarding Candidates Set (FCS) prior to data transmission. The timer-based coordination mechanism incurs the least overhead to coordinate among the FCS that has successfully received the data packet for relaying the data in a multi-hop manner. This timer-based mechanism suffers from duplicate transmissions if the FCS is either not carefully selected or coordinated. The focus of this work is to propose a hybrid opportunistic energy efficient routing design for large scale, low power and lossy WSN. This design avoids periodic 'hello' beacons for NM, limits constant overhearing and increase in packet header length. There are two modes of operation i) opportunistic ii) unicast mode. The sender node adopts opportunistic forwarding for its initial data packet transmission and instead of pre-computing the FCS, it is dynamically computed in a completely distributed manner. The eligible nodes to be part of FCS will be neighbour nodes at lower corona level than the sender with respect to the sink and remaining energy above the minimum threshold. The nodes part of FCS based on cross-layered multi-metrics and fuzzy decision logic determines its priority level to compute Dynamic Holding Delay (DHD) for effective timer coordination. The differentiated back off implementation along with DHD enables the higher priority candidate that had received data packet to forward the packet first and facilitates others to cancel its timer upon overhearing. The sender node switches to unicast mode of forwarding for successive transmissions by choosing the forwarding node with maximum trust value as it denotes the stability of the temporally varying link with respect to the forwarder. The sender node will revert to opportunistic mode to increase transmission reliability in case of link-level transmission error or no trustworthy forwarders. Simulation results in NS2 show significant increase in Packet Delivery Ratio (PDR),decrease in both average energy consumption per node and Normalized Energy Consumption (NEC) per packet in comparison with existing protocols.*

## 1. Introduction

The low power, lossy WSN consists of tiny embedded resource-constrained sensor motes limited in energy (battery-powered), bandwidth, memory and computational power [1]. The problem of time-varying channel characteristics in thewireless medium are because the radio signal propagation is subjected to large scale fading caused by reflection, scattering, diffraction and small scale fading caused by multi-path signals, Doppler effect or interferences. This results in fluctuations in signal strength and intermittent link connectivity [2].

International Journal of Computer Networks & Communications (IJCNC) Vol.11, No.6, November 2019

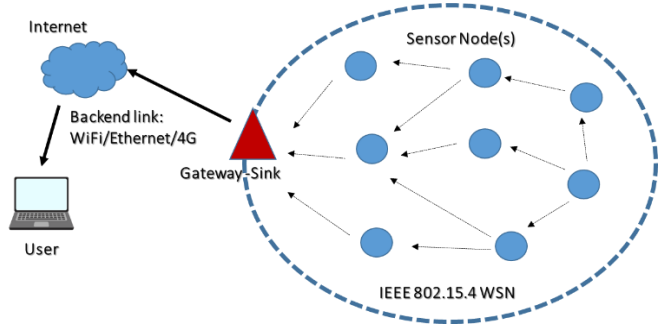

Figure 1. IEEE 802.15.4 based WSN

These unreliable wireless links make packet forwarding a challenging task and topology dynamic, despite sensor motes being stationary. Figure 1 displays the deployed IEEE 802.15.4 compliant sensor motes[3] and the sink, which acts as the gateway for network initialization, interest dissemination, data collection and external network interface connectivity. Multi-hop routing provides significant energy savings compared to direct transmission [4] and adopted in low power, lossy WSN to minimize energy cost due to communication.

The neighbouring nodes within communication range can overhear any transmission in wireless medium, since this medium is broadcast in nature. This property is utilized in OR approach and it exploits the spatial and temporal characteristics of wireless networks. For every transmission, the forwarder is chosen only after data is received by at least one in the FCS. The benefit of exploiting the broadcast nature of a wireless medium towards improving the connectivity with unreliable links has been researched in OR approaches [5 – 7].

The proposed work overcomes the limitations of existing designs with the following major contributions.

1) A Hybrid Opportunistic Routing design to achieve scalability and energy efficiency.
2) A beaconless approach by sensor motes with no periodic hello packets and routing table exchanges.
3) A corona dissemination mechanism with minimal overhead is proposed that involves no location computation via heavy overhead Global Positioning System (GPS) or having assumptions of location awareness unlike geographical routing protocols.
4) No further increase in data-packet header length, unlike traditional OR approaches.
5) A multi-metric cross-layered fuzzy decision logic for forwarding candidates to compute its priority level and DHD in a completely distributed manner. The metrics include Link Quality Indicator (LQI), trust degree of the forwarder, corona level and residual energy for making an energy-efficient forwarding choice via the opportunistic mode of transmission.
6) This Hybrid design switch to the unicast mode for successive data transmissions and minimizes problems due to constant overhearing, duplicate transmissions in WSN. However, if a link-level transmission erroroccurs or no trustworthy forwarders exists for the senderwill switch to opportunistic mode.

Section 2 discusses the related work. Section 3 explains the problem statement, Section 4 elaborates on proposed routing design. Section 5 presents a mathematical analysis for modelling delivery probability in OR and energy cost modelling of Corona Interest Dissemination. Section 6 presents the performance evaluation and simulation results using Ns2. Section 7 presents the concluding remarks and further work to be carried out.

## 2. RELATED WORK

Traditional routing designs employ route discovery to determine the end to end routes. The Pre-computed end to end route for the entire batch of data transmission may not adapt to the temporal and spatial variations in the dynamic wireless environment. These variations in the lossy radio links might incur link-level retransmissions leading to energy wastage and route re-computation. Proactive routing designs [8] leads to excessive resource consumption in low power,lossy networks. The reactive routing protocols [9] might lead to latency in the route discovery phase, but falls in the pre-select candidate category by the sender. Routing Protocols for WSN could be Data-centric, Geographic Based, Hierarchical, multipath, Quality of Service (QoS) based approaches [10-14]. The OR design makes a distributed hop by hop decision for forwarding. OR postpones forwarder selection to the receiver itself after data reception, and relies on efficient coordination mechanism to dynamically choose the best next-hop node to forward data.

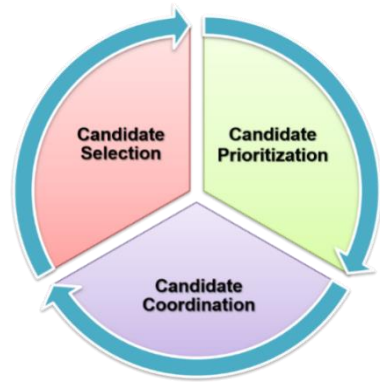

Figure 2. OR Phases

Figure2 displays the phases involved in OR. The candidate selection phase is carried out by the sender in traditional OR schemes and priority of these candidates is determined using computed routing metrics [15] by the sender. In candidate coordination phase, the FCS coordinates via either timer-based contention, token-based scheme or network coding. The pros and cons of these coordination mechanisms are elaborated in [6].

Traditional OR schemes designed for wireless networks adopts beacon-based approach. Beacon based methods transmit periodic beacon messages (hello or probe messages) to track and rank neighbours, according to computed metrics such as Expected Transmission Count (ETX) etc. ETX value can be influenced by interference among nodes, asymmetric links, multi-path fading effects etc. Existing routing metrics involve probe messages and the extra overhead involved in estimating the link quality adds significantly to the energy consumption per node. Several candidate selection algorithms focus on link state aware, delay aware and energy aware metric formulation using beacon based approaches [15], [26] – [28].

These OR schemes select and prioritize a set of candidate nodes before data packet transmission. This paves the way for the sender in OR to prioritize the list of forwarding candidates before the packet transmission. The probabilistic reception of the data packet by forwarding candidates is exploited and based on its priority order; decide when/whether to forward the received packet. This method increases the signalling overhead, and the predefined forwarding candidates list might not reflect the real situation at that instant of packet transmission. The reasons may be due to fluctuations in signal strength, environmental change impact, and malfunction in node and node's mobility.

Ex-OR (Extremely Opportunistic Routing) [5] is the first approach tested in IEEE 802.11b tested, but not energy efficient and not suitable for low power WSN. Prediction based Opportunistic Routing (POR) [16] is beacon-based OR for lossy wireless sensor network and adopts a lightweight time series prediction method for route prediction and combines link quality and geographical location to compute the priority of FCS. Beacon based OR increases the overhead and wastes resources, such as battery and bandwidth. These issues make the design of an energy-efficient and reliable routing protocol for low power, lossy wireless sensor network a nontrivial task.

In beaconless OR design, nodes need not necessarily be aware of neighbours. This saves beacon transmissions and thereby scarce resources such as bandwidth and battery-power. The nodes forward the received packets based on decision criteria and information contained in the received packet along with its state information. Link quality and Geographical beaconless routing (LinGo) [17] is one of the few works based on beaconless OR for mobile multimedia applications and employs multiple metrics for FCS selection and coordination using a timer-based contention. In Beacon-Less Routing protocol (BLR) [18], a novel idea for timer-based contention via Dynamic Forwarding Delay (DFD) computation for forwarding decisions was proposed.

Both [17-18] fall under the geographical routing category and the major assumption is nodes are aware of their locations and sink's location. The energy consumption for location estimation is ignored in these geographical OR approaches. The legacy Geographical Perimeter State Routing (GPSR) [19] does not fall under the beaconless category, but handles the void problem in case of no forwarding neighbours. In [20], the authors focus on data driven route-learning mechanism via multiple mobile sinks, but constant overhearing puts nodes in always ON state.

[21-23] focuses on real time routing protocol designs for WSN. Real-Time routing protocol with Load Distribution (RTLD) [21] restricts the forwarding area to a quadrant based area w.r.to sink and uses a multi-metric objective function for choosing the best next hop. However, this approach is not opportunistic but location based with suitable assumptions. ERTLD [22] is an enhancement to the existing RTLD and proposed a corona based framework, a viable and energy-efficient option for coarse-grained localization. This also falls under the traditional routing category and opportunism not exploited. In [24], the author presents the pros and cons of location-based routing and corona driven routing mechanism. In Opportunistic Real-Time Routing (ORTR) [23], dynamic transmission power control for making expected progress and adaptive Back off Exponent (BE) for FCS based on priority level are implemented but fall under geographical routing approach. [29-30] focuses on fuzzy logic routing using several metrics but neither opportunistic nor beaconless in approach.

In this work, a corona aware hybrid Opportunistic Scalable and Energy-efficient Routing protocol (OPSER) is proposed. This cross-layered routing design adopts a hybrid routing approach and uses multiple metrics for achieving a reliable end to end delivery and energy efficiency.

## 3. PROBLEM STATEMENT

In this work, the following issues are addressed with novel solutions.

**Issue 1:** Traditional OR schemes relyon adding a list of potential forwarders in the packetheader, whichleads to an increase in transmission cost overhead.

*Predefined FCS might not reflect the real situation at that time instant after packet transmission. Packet overhead is significantly minimized by adding only the corona level in the header instead of adding a list of "n" priority candidates.*

**Issue 2:** Existing geographic opportunistic routing approaches assume nodes are aware of its location. The overhead for location estimation is hidden in the energy cost analysis. Internet Engineering Task Force (IETF) Routing over Low Power, Lossy Networks (ROLL) working group has abandoned inclusion of GPS in nodes since it leads to detrimental behaviour on low power devices [31].

*A simplified relative location-based mechanism is used, where the relative distance of the source node to sink can be computed based on the corona level it resides ($n^{th}$ level * corona radius "R"). The corona framework with minimal overhead is proposed in this scheme.*

**Issue 3:** Traditional Distance Vector / Link State Routing Principles rely on routing table exchange on a periodic or triggered basis, which adds significant communication overhead in low power, lossy WSN.

*A hybrid routing approach is proposed based on the trustworthiness of forwarding candidates to switch between opportunistic and unicast mode of operation. This involves no exchange of routing tables and also avoids the routing loop problem.*

**Issue 4:** Sensor nodes periodic hello packet exchanges for neighbourhood management, routing metrics computation adds significantly to the energy cost.

*The current status of the link is not truly reflected by the existing pre-computed metrics prior to data transfer in a multi-hop manner. The proposed hybrid opportunistic mechanism aims to minimize packet loss/error and reduces energy cost incurring factors.*

## 4. HYBRID OPSER DESIGN

Figure3 shows the algorithmic modules involved in the proposed cross-layered hybrid opportunistic scalable and energy- efficient routing design.

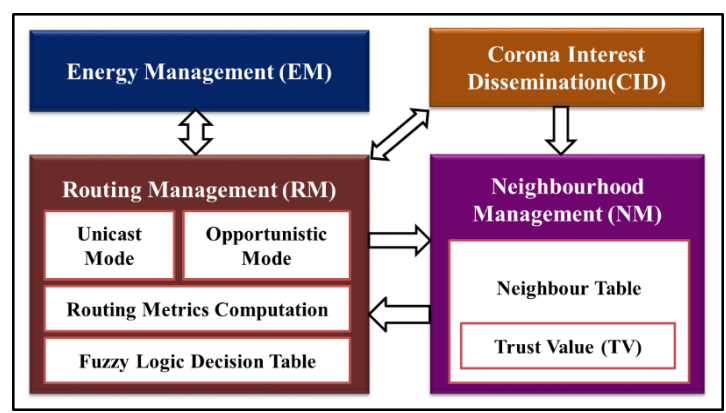

Figure 3. OPSER Modules

## 4.1. Corona Interest Dissemination (CID) Module

The corona framework divides the large-scale deployed sensor network into virtual circles centered on the sink. The corona width or radius is assumed to be of uniform width. This corona framework is a relative location-based mechanism, where the relative distance of the sensor node to sink can be computed based on the Corona Level (CL) it resides ($n^{th}$ level * corona width).

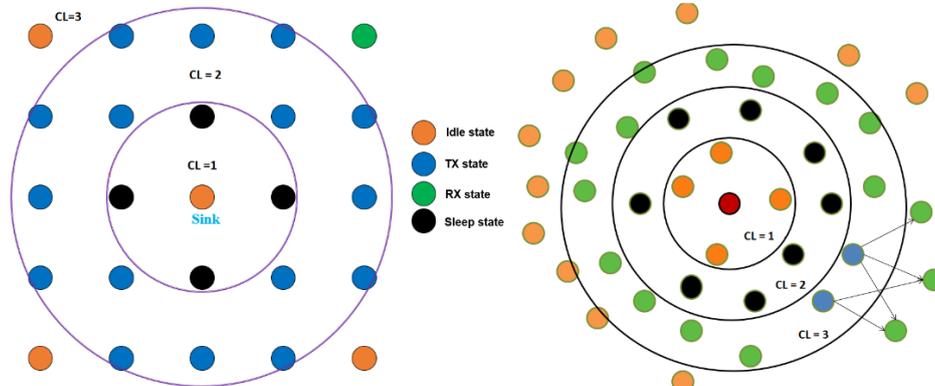

Figure 4. Corona Formation (Regular deployment) Figure 5. Corona (Random deployment)

The corona framework is similar to the tree routing but more flexible and versatile to associate with a parent or child. This mechanism allows sensor nodes to get associated with any sibling in the same level in case of a void or interim link connectivity or node failure/mobility issues [24]. This corona framework could be linked to the interest dissemination by the sink and reduce further overhead for initialization. Level based epoch scheduling for corona dissemination could reduce contention in IEEE 802.15.4 non-beaconed mode of configuration. This epoch based scheduling allows nodes in the previous level to turn to sleep state once they complete their transmission and avoid the implosion of redundant corona packets. Figure 4 and 5 shows the deployment of nodes and corona formation. The nodes switching the radio transceiver to sleep state after CID transmission to avoid redundant packet reception is also shown. Algorithm 1 explains the CID initiated by the sink and nodes learning its CL.

| **Algorithm 1 CID Module** |
|---|
| 1: function **CORONAINTERESTDISSEMINATION** |
| 2: Sink sets it's CL = 1 |
| 3: Sink adds the following header field values in CIDpacket |
|     • CID_SOURCE_ID (Sink's ID) |
|     • CID_SEQ_NUMBER (Unique Interest ID) |
|     • CL (Corona Level) |
|     • PREV_HOP_ID |
|     • NEXT_HOP_ID (=MAC BROADCAST) |
|     • CID_TTL (Time to Live of CID) |
| 4:   **for all** Sensor Node (SN) that receive CID do |
| 5:  **if** CID_SEQ_NUMBER is new **then** |
| 6:    CL of SN = CL of CID+1 |
| 7:    **if** CID_TTL is 1 **then** Drop CID |
| 8:    else CID_TTL = CID_TTL-1 |
| 9:  **end if** |
| 10: **if** (Neighbour Table (NT) is empty) OR (NEXT_HOP_ID differs) **then** |
| 11:    set SN's ROUTE_STATUS = ACTIVE |
| 12: Add a new entry in the NT with following field values set from CID packet |

```
         • SEQ_NUM
         • DESTINATION_ID
         • NEXT_HOP_ID
         • LQI derived as shown in Eq 2
         • CORONA_LEVEL
         • TRUST_VALUE = set to ½ by default
13:   SN contends to broadcast CID after updating the following field values
         • CL=CL+1
         • PREV_HOP_ID = NODE_ID (Current Node ID)
14:   Contention happens in epoch (level) based manner
15:   After transmission, the EM module switch transceiver to sleep state for $T_{sleep}$
      to avoid redundant receptions of CID ($T_{sleep} >= T_{fr} + 2T_p + T_{proc}$ units)
      Where $T_{fr}$ = Frame Transmission Time, $T_p$ = Frame Propagation delay, $T_{proc}$ = Fram
16:   processing time.
17:   end if
18:   Else
19:      Discard CID
20:   end if
21:   end for
22:   end function
```

## 4.2. Neighbourhood management (NM) Module

The design goal of the NM module is to discover a subset of forwarding candidates during the CID phase and dynamically learn, maintain neighbours on a per-packet basis to compute its Trust Value (TV) based on node's past contribution. Upon fresh learning of a forwarding neighbour, TV is initialized to 0.5 (max value $TV_{max}$ of 1.0 and min value $TV_{min}$ of 0). This threshold of 0.5 is set as the starting value as it could either become trust worthy or non-trust worthy based on the status of future data packet transmission. If the data transmission by the sender to that forwarding node is unsuccessful, the node trust value is penalized by 50 percent of its existing value to make it non-trust worthy and hence the forwarding node(s) with TV < 0.5 is classified as non-trust worthy. For every successful transmission by the forwarder, the sender increases the trust value of it by 10% and continues until it reaches the max value of 1. However, if a link-level retransmission occurs, the trust value is reduced by 50% for that forwarder and could participate only if it wins contention when switched to opportunistic mode. If it wins, the sender sets the trust value of that forwarder back to 0.5. The LQI($0 \leq LQI \leq 255$)on the reverse link from that forwarder to the sender is also updated and averaged on a per-packet basis. Table 1 shows the Neighbour Table (NT) format and sample entries.

Table 1. Neighbour Table (NT).

| Forwarder ID | Trust Value (TV) | LQI |
|---|---|---|
| 3 | 0.7 | 160 |
| 5 | 0.25 | 132 |

## 4.3. Routing Management (RM) Module

The sensor node can play the role of the sender, receiver and forwarder. The source node, which generates the packets, may need to pass on multiple levels to reach the sink. The task action handler upon generation and reception of data are programmed accordingly. The summary of the RM module functionalities is given below.

1. Source Node for its initial data packet transfer operates in opportunistic mode to set up a virtual path to reach the sink via multi-hop.
2. Data Packet is broadcast by Source Node after embedding its CL and the eligible Forwarding Candidates (FCs) after reception of data packet are determined on the fly in a completely distributed manner.
3. These FCs compute cross-layered routing metrics and determine their priority order based on the fuzzy logic decision table.
4. Based on the priority order, FCs calculates the Dynamic Holding Delay (DHD) for timer-based coordination scheme.
    1. E.g. The FC with highest priority will have the least holding delay to win contention and enable other FCs to cancel their timers upon overhearing of the same packet sent by that FC. This entire process is achieved by DHD computation and differentiated back off MAC mechanism.
    2. The algorithm continues in the opportunistic mode for the initial packet transfer tillit reaches the sink, which sends an explicit ACK.
5. The Sender Node (SN) uses the transmitted packet by that FC as a Passive ACK to transmit the subsequent packets in unicast mode tilla link-leveltransmission failure occurs for the sender node to revert to opportunistic mode.

### 4.3.1. Send Data Packet function

Algorithm 2 explains the role of Send Data Packet () function when a node acts as a source of data.

| Algorithm 2: RM Module |
|---|
| 1: function **SENDDATAPACKET** |
| 2: **if** (PACKET_ID == 1) \|\| LookupNTMax(TV) < $\frac{1}{2}$) \|\| (ROUTE_STATUS == FAILED) **then** |
| 3: Source Node (S) in Opportunistic mode |
| 4: Add following fields in Data Packet Header |
|     • CL (S's Corona Level) |
|     • PACKET_ID (Unique Identifier) |
|     • DESTINATION_ID (Sink ID) |
|     • SOURCE_ID (S's ID) |
| 5: Contend and do MAC_BROADCAST of the data packet |
| 6: Wait for Passive ACK (overhearing the data packet forwarded by neighbour) within $DHD_{max}$ |
| 7: **if** (overheard same data packet rebroadcast) **then** |
| 8:     **if** (NT == Empty) \|\| (Neighbour who rebroadcasted doesn't exist in NT) **then** |
| 9:     Add that neighbour in NT with TV = $\frac{1}{2}$ |
| 10:     **else** |
| 11:     Increase the neighbour's TV by 10%  ($TV_{max} = 1$) |
| 12:     **end if** |
| 13:   **if** S's ROUTE_STATUS == FAILED **then** |
| 14:   set ROUTE_STATUS = ACTIVE |
| 15:   **end if** |
| 16:   **else** (Timeout) |

```
17:        MAC_LEVEL_RETRY
18:      end if
19: end if
20: if LookupNTMax (TV≥ 1/2 ) then
21:      S in Unicast mode
22:      Choose forwarding neighbour with max TV and LQI >= LQI_{TL}
23:      Unicast Data packet to the chosen forwarding neighbour
24:      if (ACK) then
25:         Increase that neighbour's TV by 10%
26:         TV of the chosen neighbour in NT =TV + 10%*TV (TV_{max}=1)
27:      else
28:         Decrease that neighbour's TV by 50%
29:         TV of the chosen neighbour in NT = TV - 50% *TV (TV_{min}=1)
30:         Set S's ROUTE_STATUS = FAILED
31:         Revert to opportunistic mode
32:      end if
33: end if
34: end function
```

### 4.3.2. Recv Data Packet function

Algorithm 3 explains the role of Recv Data Packet () function when a node receives data packet.

**Algorithm 3: RM Module**
```
 1: function RECVDATAPACKET
 2: NB: Number of Backoff
 3: BE: Backoff Exponent
 4: macMinBE: Minimum BE as per IEEE 802.15.4 MACspecification
 5: macMaxBE: Maximum BE as per IEEE 802.15.4 MAC specification
 6: DHD: Dynamic Holding Delay
 7: NB=0
 8: LQI_{norm} = Calculate from Instantaneous LQI of the received packet as per Eq 4
 9: E_{rem} = Obtain the remaining power level of the node
10: deg_{trust} = Number of trustworthy forwarding neighbours
11: Sensor Node (SN) has received data packet
12: if (SN's CL > CL in the received packet) then
13:    Drop the data packet
14:  if (SN's CL <= CL in the received packet)&& (E_{rem}≥ E_{min})) then
15:    if (DataPkt is MAC_BROADCAST) then
16: Compute DHD based on Fuzzy Logic Table Lookup(LQI_{norm},deg_{trust}) as per Table 2
17: if (Packet with same SEQ_NUMBER overheard before DHD expires) then
18:       Abort the transmission
19:     end if
20:     set BE = macMinBE according to its priority level
21:     Delay for random (2^{BE} - 1) unit backoff periods
22:     Perform CCA on backoff period boundary
23:     if (Channel Idle) then
24:       forwardDataPacket()
25:     else
26:       NB=NB+1
27:       BE=min(BE+1,macMaxBE)
```

```
28:     end if
29: if (NB >macMaxCSMABackoffs) || (overheard same SEQ_NUMBER packet) then
30:      Drop the Packet
31:     end if
32:   end if
33:   if (DataPkt is UNICAST) then
34:     Contend to forward
35:     if (Channel Idle) then
36:       forwardDataPacket()
37:     else
38:       as per Binary Exponential Backoff(BEB) algorithm
39:     end if
40:   end if
41:  end if
42: end if
43: end function
```

### 4.3.3. Forward Data Packet function

The Forward Data Packet function is similar to the send Data Packet function except that it forwards the packet received from a sender node in multi-hop to potential lower corona level nodes till it reaches the sink when it is part of the forwarder's NT. The forwarder operates in the opportunistic mode of forwarding or unicast mode of forwarding similar to the send Data Packet function.

### 4.3.4. Metrics computation and Fuzzy Decision Logic for FCS

During the opportunistic mode, OPSER postpones the decision of candidate selection and forwarding to the receiver side after data transmission. Unlike traditional beacon-based OR, the priority level is not determined by the sender in this design and is computed by the FCS in a completely distributed manner. The challenges involved to achieve this goal are to ensure that

- Nodes part of FCS should be designed to avoid choosing the same back-off periods as it leads to high channel contention and collision. Hence, the necessity arises for the adaptive back-off mechanism depending on the priority level.
- In case of a tie in priority level for nodes in FCS, the design should attempt to minimize collision, duplicate transmission and cancel the timer upon overhearing by another forwarder having the same priority level.

Table 2. Multi-metric fuzzy decision logic for FCS

| $LQI_{norm}$ | $deg_{trust}$ | Priority Level | macMinBE | macMaxBE |
|---|---|---|---|---|
| HIGH | HIGH | 1 | 2 | 4 |
| HIGH | LOW | 2 | 3 | 5 |
| MED | HIGH | 3 | 4 | 6 |
| MED | LOW | 4 | 5 | 7 |
| LOW | HIGH | 5 | 6 | 8 |
| LOW | LOW | 6 | 7 | 9 |

The potential forwarders belonging to the FCS compute their priority level based on the normalized LQI of the received packet, trust degree of the forwarder in terms of the number of trustworthy forwarding candidates. The fuzzification involves the conversion of these crisp inputs into a fuzzy set. Table 2 shows the multi-metric fuzzy decision logic table. The participating nodes for contention are restricted to only forwarder nodes operating at least beyond the minimum threshold level ($E_{min}$) of its remaining energy ($E_{rem}$). The holding time i.e Dynamic Holding Delay (DHD) is calculated based on Equation 1.

This multi-metric fuzzy decision logic increases robustness, reliability in data transfer and facilitates to achieve energy efficiency. To address the void problem in case of no forwarding candidates, the nodes at the same level would have the lowest priority level assigned with value "7" to participate if no lower-level nodes exist.

The defuzzification step involves determining the priority level for nodes in FCS i.e. crisp output based on the match in the fuzzy decision logic table.

$$\text{DHD}_{\text{FCS}_i} = (\text{Priority Level} - 1) * \text{T} + \tau \qquad (1)$$

Where, $\text{DHD}_{\text{FCS}_i}$ is the computed DHD value of $i^{th}$ candidate in the FCS ($i \in \deg_{\text{SN}_{\text{FCS}}}$),

$\deg_{\text{SN}_{\text{FCS}}}$ is the number of candidates in FCS of Sender Node (SN),
T is the predefined holding time,
$\tau$ is the random delay factor between $[\tau_{min}, \tau_{max}]$ such that
$\text{DHD}_j < \text{DHD}_{j+1}$, j and j+1 are the priority level of the candidates in FCS.

### 4.3.5. Link Quality Indicator Metric

The sensor node's physical layer in IEEE 802.15.4 compliant radio chip is capable of estimating the instantaneous LQI and RSSI for every received packet. Due to the asymmetric and dynamic nature of wireless links, it is necessary to track the instantaneous LQI for achieving a higher packet delivery ratio (PDR). LQI can be modeled as per [32].

$$\text{LQI} = 255 * \frac{\text{RSSI} - \text{ED}_{min}}{\text{ED}_{max} - \text{ED}_{min}} \qquad (2)$$

Where,

- LQI is the instantaneous Link Quality Indicator of the received packet (8 - bit unsigned integer)
- RSSI is instantaneous Received Signal Strength indicator of the received packet in dBm,
- ED is Energy Detection (minimum and maximum) values in dBm.

RSSI can be modeled using the log-normal shadowing signal propagation model [33].

$$\text{RSSI} = P_t - \text{PL}(d_0) - 10\beta \log \frac{d}{d_0} + X_{\sigma_{dB}} \qquad (3)$$

- $P_t$ is the transmit power level in dBm,
- d is the separation distance between the sender and receiver part of FCS,
- $\text{PL}(d_0)$ is the path loss at reference distance $d_0$,
- $\beta$ is the path-loss exponent,
- $X_{\sigma_{dB}}$ is a Gaussian random variable with zero mean and standard deviation $\sigma_{dB}$,
- $\sigma_{dB}$ is the variations of received signal power due to multipath fading.

The fuzzy levels for $LQI_{norm}$ is based on the following equation

$$LQI_{norm} = \begin{cases} LOW & LQI_{(S,FC_i)} \leq LQI_{TL} \\ MED & LQI_{TL} < LQI_{(S,FC_i)} < LQI_{TH} \\ HIGH & LQI_{(S,FC_i)} \geq LQI_{TH} \end{cases} \qquad (4)$$

- $LQI_{norm}$ is the normalized LQI
- $LQI_{(S,FC_i)}$ is the instantaneousLQI of received packet between the sender node and i[th]candidate.
- $LQI_{TL}$ is the lower threshold of LQI
- $LQI_{TH}$ is the higher threshold of LQI

### 4.3.6. Trust degree Metric

Let $i_{deg_{FCS}}$ represent the degree of forwarder i having trustworthy forwarding candidates of atleast 0.5. The trust degree of the forwarder is mapped as

$$deg_{trust} = \begin{cases} HIGH & i_{deg_{FCS}} > 2 \\ LOW & 1 \leq i_{deg_{FCS}} \leq 2 \end{cases} \quad (5)$$

The nodes in FCS with $i_{deg_{FCS}} > 0$ will only participate in timer-based contention.

### 4.3.7. Adaptive Backoff Exponent (BE) parameters

To facilitate cancellation of scheduled transmission upon overhearing of the same packet by another priority forwarder and to minimize high channel contention, the adaptive back off mechanism is proposed. As shown in Table 2, the highest priority node is assigned a shorter BE to provide a higher chance to first transmit the data. The forwarder in FCS initializes BE to macMinBE value according to its priority level in Table 2. A back off period is then a random number between $0 to 2^{BE} - 1$. Once the back off period is completed, it performs a clear channel assessment (CCA). If the channel is clear, the node transmits. Else if busy, BE is incremented tillmacMaxBE and repeats the process till Number of Backoffs (NB) times to declare in the worst case as Channel Access Failure (CAF). If the forwarder receives the same packet that it tries to send, the packet is discarded.

### 4.4. Energy management (EM) Module

The main function of the EM module is to select the discrete transmission power level of the radio transceiver in the sensor node. The state of the transceiver is possibly in transmit, receive, idle and sleep state. The nodes that are not part of FCS can revert to sleep statetill next active cycle. Also during the corona dissemination phase, once the node had completed its CID packet transmission after learning its corona level can switch to the sleep state till next active cycle. To enhance the network life span, nodes having remaining energy ($E_{rem}$) beyond the operating threshold ($E_{min}$) only take part in FCS.

## 5. MATHEMATICAL ANALYSIS

### 5.1. Network Model

Let G(V,E) represent the dynamic graph modelled as Destination Oriented Directed Acyclic Graph (DODAG) where
- $V_i$ represents the sensor node,$i \epsilon N$, N is the number of nodes in the network,
- E represents the time-variant edge connectivity between two adjacent nodes,
- the edge weight w(i,j) also depends on the spatial and temporal variations due to channel fluctuations,
- the root node (D) is the sink (Destination),
- the depth of the sensor node $V_i$ is determined in-terms of Corona Level (CL) from the root node and represents the number of edges in a unique directed path from D to $V_i$.

Each sensor node $V_i$ has a battery with available energy ($E_{initial}$) at its initial stage. The $V_i$'s radio transceiver is configured to operate a fixed transmit power level $P_t$.

## 5.2. Energy Cost Modelling of CID Algorithm

The total energy cost involved in the CID in the network is represented as follows

- Let $E_{tx}$ be the transmission cost involved in the transmission of CID Packet.
- Let $E_{rx}$ be the reception cost for receiving the CID packet.

The reception cost for every node at level 'i' includes only nodes from level 'i-1' and not from 'i+1', as the radio transceiver in level 'i' is put to the sleep state after it completes its CID transmission to avoid reception from level 'i+1'. Every node in the network at level 'i' transmits CIP only once upon reception of the CIP packet from previous level 'i-1' nodes.

Per node ($V_i$)'s energy cost involved in disseminating CIP packet

$$= E_{tx} + deg_s(V_i) * E_{rx} \qquad (6)$$

Where, $deg_s(V_i)$ represents the sub-set degree of $V_i$.

Total Energy Cost per node involved during the entire lifetime of an application

$$= \sum_{i=1}^{t} E_{tx} + deg_s(V_i) * E_{rx} \qquad (7)$$

Where, t is the number of rounds of CID dissemination.

In this work, t is set to 1 for network initialization and interest dissemination by the sink.

For multi-hop network deployment and node having dense neighbour set,

$$deg_s(V_i) \approx \log(N) \qquad (8)$$

Total energy cost involved in disseminating CID throughout the network

$$= N * E_{tx} + N * deg_s(V_i) * E_{rx} \qquad (9)$$

Where, $N$ is the number of nodes in the network.

In the worst case, the total energy cost involved in disseminating CID packet in the network will be $N * E_{tx} + N * \log(N) * E_{rx} \approx O(N\log N)$ (10)

## 5.3. End to End delivery probability analysis

Figure 6 shows the directed graph in which the link delivery probability from source to forwarding candidates are known. The packet has to travel N hops to reach the sink.
In case of opportunistic forwarding, the probability that the transmission from Source S reaches the root sink D via 'N' number of hops is

$$P_O = \left(1 - \left(1 - P_{FC_1}\right) * \left(1 - P_{FC_2}\right) * \left(1 - P_{FC_3}\right) \ldots * \left(1 - P_{FC_{S_{deg(FCS)}}}\right)\right)^N \qquad (11)$$

Where, $P_{FC_1}, P_{FC_2}, .., P_{FC_{S_{deg(FCS)}}}$ represents the link delivery probability of S to each of the FCS nodes and $P_{FC_i} \neq 1$.

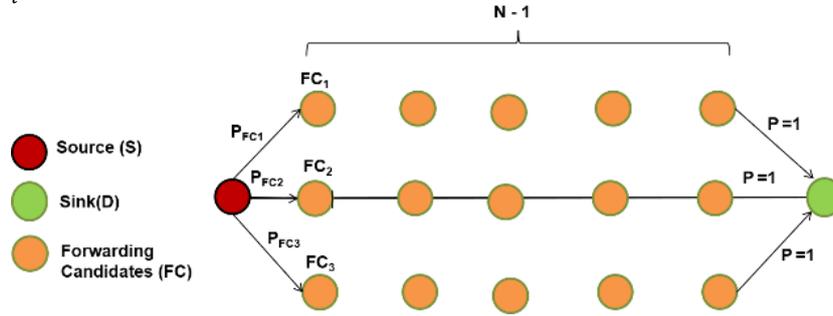

Figure 6. Delivery probability Analysis –End to End

In case of unicast forwarding, the probability that the transmission from Source S reaches the root sink D via 'N' number of hops by choosing the best forwarding candidate in every hop is

$$P_U = \left(1 - \left(1 - P_{FC_i}\right)\right)^N \qquad (12)$$

Hence, Opportunistic forwarding has a higher delivery probability since $P_O > P_U$

Purely Opportunistic Mode of Forwarding improves the transmission reliability as at-least one of its FC would probabilistically receive the packet. The end to end delivery rate could thereby be improved. However, the timer-based coordination might incur duplicate transmissions (due to hidden terminals) and energy cost incurred needs to be minimized. Also, nodes in constant overhearing mode could waste more energy. In Purely unicast driven routing, the link delivery probability should be utilized to the fullest and link-level retransmissions needs to be reduced. OPSER adopts a hybrid approach to improve end to end delivery rate and reduce energy cost incurred.

## 6. PERFORMANCE EVALUATION

### 6.1. Simulation Parameters & Configuration

NS2 [34], a discrete event simulator is used for the implementation of the proposed protocol and performance evaluation. Table 3 shows the simulation parameters and configuration in NS2.

Table 3. Simulation Parameters.

| Network Parameter | Value |
| --- | --- |
| ***Physical Layer*** | ***IEEE 802.15.4*** |
| Data Rate ($R_b$) | 250 Kbps or 62.5 KSymbols/sec |
| Transmit Power Level $P_t$ | 0dBm (1mW) |
| Receiver Sensitivity(RxThresh_, CSThresh_) | -110dBm ($1e^{-14}$) |
| Antenna Configuration -Omnidirectional Antenna | $G_t = 1; G_r = 1; L = 1,$ $h_t, h_r = 0.03125m$ [35] |
| Log Normal shadowing (Model 1) | $\beta$=4.5 ;$\sigma$= 4dB, $d_0$ = 1m |

| Two Ray Ground Reflection (Model 2) + Error Rate (in units of packets) = | Uniform Random Variable distributed between 0.0 and 1.0 |
|---|---|
| ***MAC Layer*** | ***IEEE 802.15.4 non beaconedCSMA/CA*** |
| macMinBE, macMaxBE | Default values 3, 5(Refer Table 2 for FCS) |
| numofCSMABackoffs | 7 [25] |
| macRetries (Number of mac level retries) | Default 3 |
| Holding Time (T) | 5ms |
| ApplicationTraffic/Transport | CBR/UDP |
| Initial energy | 3.6 Joules |
| Power spentper packetfor transmission/reception | 0.02955W / 0.0255W [21-22] |
| Packet Rate & Packet size | 5 packets / sec & 70 bytes |

## 6.2. Performance metrics

The metrics computed for performance evaluation in NS2 are listed below:

**Packet Delivery Ratio (PDR)** is the ratio of packets successfully delivered to sink among the total packets transmitted by the source(s).

$$\text{PDR} = \frac{\text{Total number of received packets at sink}}{\text{Total number of sent packets by source(s)}} \quad (13)$$

**Average End to End Delay** ($D_{avgE2E}$) is the average delay taken by data packets to travel from source to sink (in seconds).

$$D_{avgE2E} = \frac{\text{Sum of received packets delay at sink}}{\text{Number of received packets at sink}} \quad (14)$$

**Total energy consumption (TEC)** is the sum of total consumed energy or used up energy by all nodes in the deployed network until the end of the simulation (in Joules).

**Average energy consumed per node ($\bar{E}$)** is defined as

$$\bar{E} = \frac{\text{Total Energy Consumed (TEC) by all nodes in the network}}{\text{Number of nodes deployed in the network}} \quad (15)$$

**Normalized energy consumption (NEC)** per packet (in Joules/packet) is defined as

$$\text{NEC} = \frac{\text{TEC}}{\text{Total number of received packets at sink}} \quad (16)$$

## 6.3. Impact of radio propagation modelling on packet reception probability

It is evident from Figure 7, that there are three regions of connectivity i.e. connected, transitional and disconnected region [33] with varying distance of separation. The probability of packet reception in the transition region exhibits the lossy link characteristics. Choosing nodes in the connected region may lead to an increase in the number of hops. The higher path loss exponent and shadowing deviation parameters in the lognormal shadowing impact the connectivity with higher variance in the received signal power modelling the lossy-link behaviour.

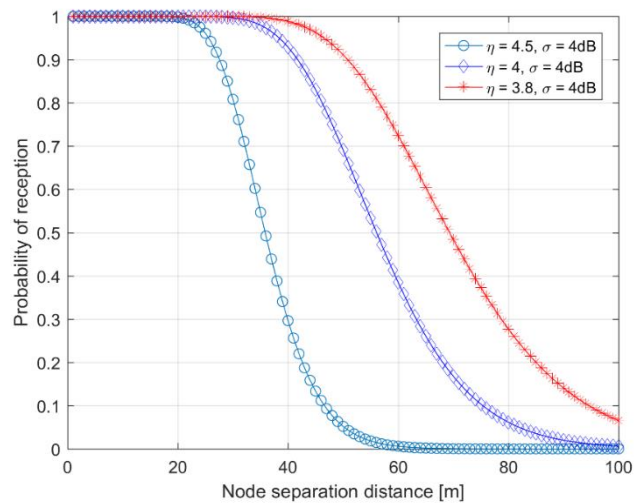

Figure 7. Impact of ($\beta$, $\sigma_{dB}$) vs separation distance on Prob. of reception

## 6.4. Impact of 'BE' on IEEE 802.15.4 non-beaconed CSMA/CA MAC performance

The number of traffic sources was set to N-1, where N is the number of nodes. The last node is the sink node. Figure 8 shows the simulation results when the number of traffic sources was increased i.e.the number of nodes were scaled up; the choice of macMinBE, macMaxBE impacts the PDR.At low traffic loads, transmission succeeds as collision probability is reduced and waiting times are also minimal. However, as the traffic load increases, the PDR drops, as more nodes contend for the channel leading to collision and transmission failure. Higher the range of random back-off period, better the delivery rate in high traffic loads. The default values of macMinBE, macMaxBE set to 3,5 as fixed values without priority differentiation in FCS will not provide fair competition and bandwidth allocation.

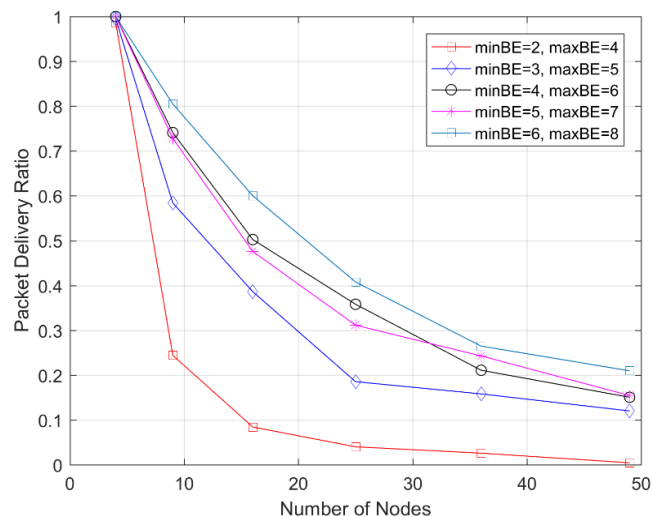

Figure 8. Impact of increase in number of nodes (Traffic Sources) vs BE parameters on PDR

## 6.5. Impact of multiple traffic sources and traffic rate

Many to one converge-cast type of traffic pattern is a common scenario in WSN, where multiple sensor nodes generate data traffic to the sink. The results presented in this section are based on the fixed network size of 121 nodes (11*11). To increase the number of hops between the source and sink, the source nodes and sink are chosen at the extreme ends of the deployment. The number of traffic sources is set to 4 during the simulation period. OPSER is compared with real time protocols such as RTLD [21] that consider multi-metrics such as Packet Reception Rate (PRR), packet velocity and residual energy for forwarder selection. Real-Time Power Control (RTPC) [36] forwards data packets by choosing the best energy efficient forwarder using packet velocity as the metric to minimize hop count. Prediction based Opportunistic routing (POR)is investigated under IEEE 802.15.4 in NS2.

### 6.5.1. PDR vs Traffic Rate

Figure 9 shows the PDR when the packet rate is increased by multiple traffic sources. This leads to high channel contention for delivery of data by sensor nodes. It is evident that with multiple traffic sources and error model, the protocols suffer in PDR with an increase in packet rate. OPSER shows high PDR performance and significant improvement compared to real time (RTLD, RTPC), reactive (AODV) and opportunistic (POR) routing protocols. The opportunistic mode of transfer and best forwarder choice in terms of link quality, trust degree of the forwarder has increased the reliability in delivery of data towards the sink. The adaptive back off exponent for high priority forwarder and timer cancellation by other FCS candidates reduces high channel contention. Switching to the unicast mode for successive transmissions to most trustworthy forwarders has reduced the number of nodes contending for a channel; thereby minimize CAF and probability of collision.

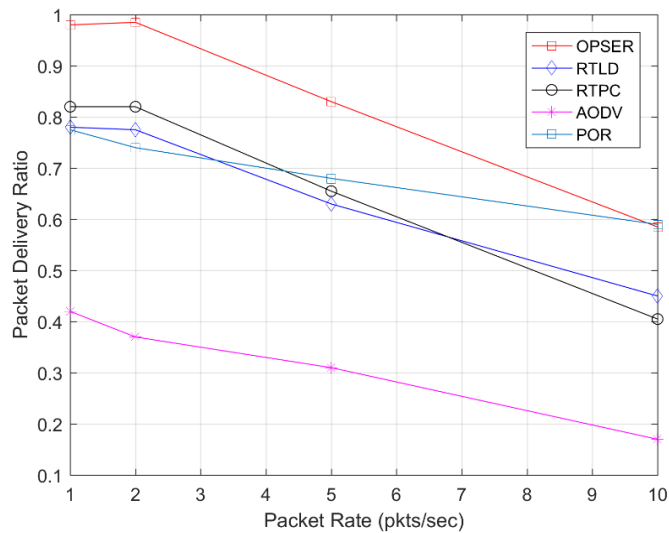

Figure 9. PDR vs Packet Rate (p/sec)

Though RTLD focuses on multi-metric such as PRR, packet velocity and residual energy for forwarder selection, it is not opportunistic to combat the lossy links. The amount of route requests and route replies sent by sender and forwarders also leads to more channel contention and the possibility of collisions.

### 6.5.2. Average Energy Consumed vs Traffic Rate

Figure 10 shows the average energy consumed per node in the network when packet rate is increased for fixed network size. It is evident that there has been a significant reduction in the average energy consumed per node as several overhead factors such as periodic hello beacons, route requests, replies are avoided in OPSER. The maximum number of packets generated by the traffic sources is fixed in the simulation. However, with the increase in packet rate by these sources and lossy links, the real time routing protocols such as RTLD, RTPC experience higher energy consumption due to route request, reply by sender and forwarders incurring extra

International Journal of Computer Networks & Communications (IJCNC) Vol.11, No.6, November 2019

overhead. POR incurs more overhead with periodic hello packets leading to higher energy consumption and channel contention. OPSER minimizes the average energy consumed per node relatively when compared to reactive AODV protocol primarily because of avoiding route request flooding and theroute replies. Also, opportunistic forwarding in terms of multiple metrics avoids retransmission and achieves energy efficiency.

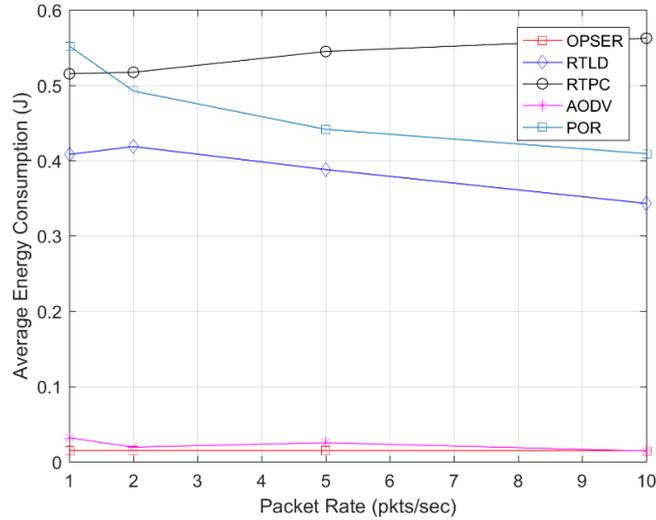

Figure 10. Average Energy Consumed (J) vs Packet Rate (p/sec)

## 6.6. Impact of scaling the number of nodes in the network

The results in this section are based on the performance of the protocols under the lognormal shadowing model as per Table 3. The source and sink is chosen at the extreme opposite diagonal ends to increase the number of hops and forwarding region. The nodes are distributed in a regular grid manner with a uniform spacing of 10m between nodes.

### 6.6.1. PDR vs. Number of Nodes

Figure 11 shows the PDR performance by varying the number of nodes in the simulation.

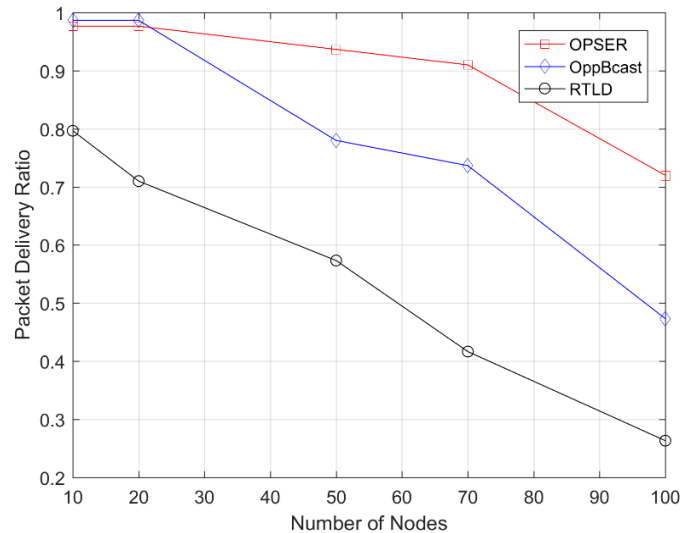

Figure 11. PDR vs Number of Nodes

It is evident that RTLD degrades in packet delivery ratio with an increase in the number of nodes and lossy link characteristics model.Due to lossy links and not adapting to the temporal variations, the PDR degrades as the number of hops scales up in real-time routing protocols. The fully Opportunistic Broadcast (OppBcast) is a simplistic randomized timer contention mechanism and limited to only nodes part of FCS. Since FCS is not metric based in OppBcast, it is not an efficient forwarding choice leading to lesser delivery ratio. In addition, due to poor timer coordination because of randomized contention may leads to duplicate transmission and hidden terminal issues. Whereas in OPSER, priority based timer contention during opportunistic mode of transfer and best forwarder choice in terms of link quality, trust degree of forwarder has increased the reliable data transfer to sink. Switching to the unicast mode for successive transmissions to most trustworthy forwarders reduces more nodes contending for the channel.

**6.6.2. Normalized Energy Consumption vs Number of Nodes**

Figure 12 shows the normalized energy consumption per packet when the number of nodes is increased. It is clearly evident that OPSER achieves the least in terms of normalized energy consumption as the overhead is significantly reduced and the best energy efficient forwarding choice via opportunistic mode combats the lossy links. The real time routing protocol RTLD experiences higher energy consumption due to route request, replies incurring extra overhead and not truly reflecting the current status of lossy links as it's not opportunistic. OPSER minimizes the normalized energy consumption per packet and achieves energy savings relatively when compared to OppBcast because of the multi-metric fuzzy logic decision by forwarding candidates.

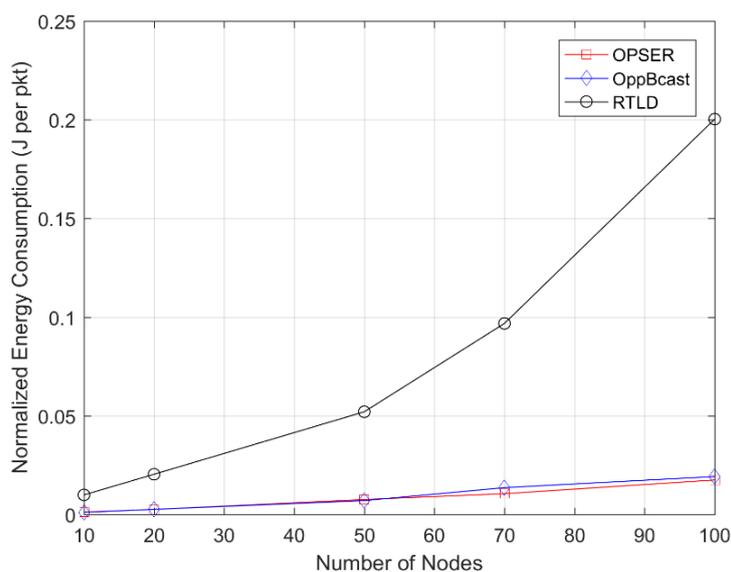

Figure 12. Normalized energy consumption (J/P) vs Number of Nodes

**6.6.3. Average End to End Delay vs Number of Nodes**

Figure 13 shows the average end to end delay with an increase in the number of nodes. The trade-off factor is visible where OPSER has increased average end to end delay performance in comparison with RTLD. OppBcast has comparatively least average end to end delay as it operates fully in opportunistic mode and minimizes the time to deliver the packet without priority based differentiation among FCS. The increase in average end to end delay in OPSER is because of DHD computation, the additional holding time for priority differentiation.

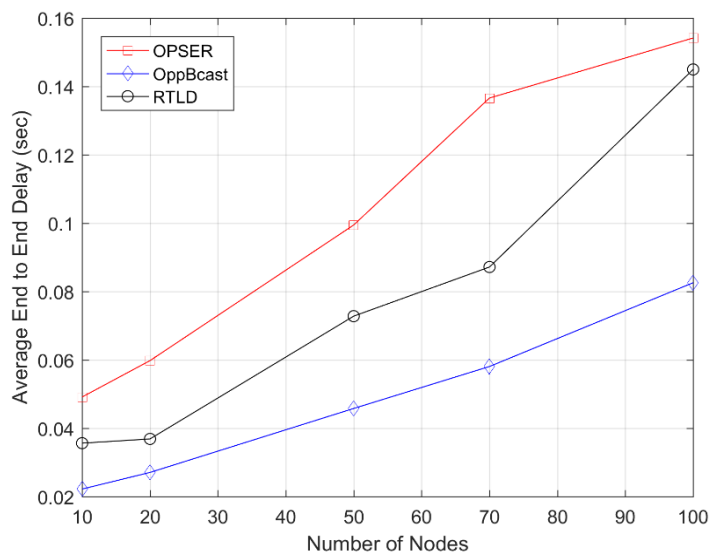

Figure 13. Average End to End Delay (sec) vs Number of Nodes

However, this trade-off factor of additional end to end delay comes with significant energy savings. Hence, OPSER design can potentially suit for large scalenon time-critical applications.

## 7. CONCLUSIONS

In this work, we proposed an OPSER algorithm for lossy WSN that uses a hybrid approach based on the trustworthiness of forwarding candidate to switch between opportunistic and unicast modes of operation. The proposed protocol is found suitable for large-scale deployment despite problems such as lossy links and constrained sensor nodes with limited available energy. Simulation results validate that OPSER algorithm outperforms existing algorithms in terms of PDR, average energy consumed per node and normalized energy consumption per packet. OPSER minimized packet loss since, during the opportunistic mode of transfer, forwarding candidates compute its priority level in a completely distributed manner based on cross-layered multi-metric fuzzy decision logic. The metrics included link quality, trust degree of the forwarder, remaining energy and corona level for making an energy-efficient forwarding choice to increase the reliability in data transmission for the multi-hop network.

The simulation in this work is limited to static deployment of nodes and further would be extended to dynamic topology due to node mobility. The sink would be quasi-mobile and sensor nodes moving out of corona level would reset its level information and perform dynamic learning of the corona level with minimal overhead.

**AUTHORS**

**Jayavignesh Thyagarajan** obtained Bachelors in Engineering (B.E) degree in Electronics and Communication from College of Engineering, Guindy, Chennai, India in 2003.He acquired his Masters in Engineering (M.E) from Madras Institute of Technology, Anna University, India and secured First Rank with a Gold medal in Communication and Networking program in 2010.He had served as Systems Engineer in Tata Consultancy Services (TCS), India in Telecommunication sector during 2010 to 2013. He is currently working as Assistant Professor in the School of Electronics Engineering (SENSE) in Vellore Institute of Technology (VIT), Chennai Campus, India. He is pursuing PhD and his research interests includes cross layered routing design in the area of Wireless Adhoc, Sensor, Mesh, Vehicular and Opportunistic networks.

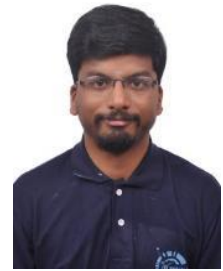

**Suganthi K** acquired Bachelors in Engineering (B.E) degree in Computer Science and Engineering from Madras University in 2001, Masters in Technology (M.Tech) in Systems Engineering and Operations Research and PhD in Wireless Sensor Network from Anna University in 2006 and 2016 respectively. She is currently working as Assistant Professor (Senior) in the School of Electronics Engineering (SENSE) at Vellore Institute of Technology (VIT), Chennai campus, India since 2016. Her research interests include Wireless Sensor Network, Internet of Things, Data Analytics and Network Security.

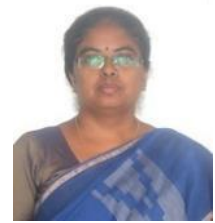